# Fluorescence temperature sensing based on thermally activated singlet-triplet intersystem crossing in crystalline anthracene

CHENG TANG,[1] XIAOJUN ZHU,[1] YUNFEI SONG,[2] WEILONG LIU,[1] QINGXIN YANG,[1] ZHE LV[1] AND YANQIANG YANG[1,2,*]

[1]*Department of Physics, Harbin Institute of Technology, Harbin, 150001, China*
[2]*National Key Laboratory of Shock Wave and Detonation Physics, Institute of Fluid Physics, China Academy of Engineering Physics, Mianyang, 622900, China*
*\*yqyang@hit.edu.cn*

**Abstract:** The temperature dependence of the steady-state fluorescence spectrum of anthracene crystals range from 300K to 500K had been investigated, which was in the temperature range of most tabletop laser-driven shock wave experiments. The interesting finding is that the fluorescence intensity of the 2-0 transition increases more rapidly than other transitions with the rising temperature. In particular, the transition intensity ratios $γ_n$ all shows a perfect exponential increasing curve, which can be used for fluorescence temperature sensing. The analysis of sensitivity $η$ and random uncertainty $ΔT$ has demonstrated that the intensity ratio $γ_1$ is the best comprehensive performance physical quantity for temperature sensing. The theoretical analysis and experimental results demonstrated that unusual intensity increasing of 2-0 transition was originated from the second excited triplet state $T_2$, which was thermally coupled with the first excited singlet sate $S_1$. In a word, we established a new fluorescence temperature sensing method based on the intensity ratio and clarified the mechanism of this method was the thermally activated singlet-triplet intersystem crossing.

## 1. Introduction

Note the ability to provide microscopic insights into the response of condensed matter under extreme and unique conditions, such as dynamite explosion [1], earthquake [2] and meteorite impacts [3], the research of shock wave has been continuing to attract considerable attention from academia for over a century. Detailed understanding the pressure and temperature of the materials under shock compression has been a hot topic in material researches [4, 5]. In particular, it is essential to derive the equation of state (EOS) for modulating the behavior of material under shock wave compression [6, 7]. Due to the advantage of contactless measurement, time resolved Raman spectroscopy has been widely used as an efficient diagnostic methodology for investigating the pressure distribution in shocked materials [8-11]. However, note the temperature induced Raman mode frequency shift was negligible, the effect of temperature was been routinely ignored in those researches [9, 10]. Hence, the temperature sensing in the shocked material is still an intense and significant research field in recent decades.

The research of optical temperature sensing has been last for a long time. Note that the infrared imaging and Raman thermography was limited by the thermo-optical properties of sample, the fluorescent temperature sensing has attracted most attention [12, 13]. With the advantages of the fast response in a time scale of cellular process and the high spatial resolution, the organic molecular probes were widely used for the researches of chemistry and biology [14, 15]. A typical kind of organic molecular probe is the Rhodamine and its derivatives. Such as, the Rhodamine 6G molecule has been used as a tracer for recording the conformation change of DNA at different temperature [16]; the Rhodamine B was widely used in the temperature measurement in microfluidic systems because of the cheap price and the broad working range [17, 18]. Albeit there were lot of reported researches of the fluorescence spectrum of anthracene [19-21], due to the complex fluorescent components and the broad and structureless excimer emission, the fluorescent temperature sensing possibility of anthracene has not been explored. Gupta *et al.* found that the excimer emission would not appear in the anthracene crystals during the pressure loading process [22], which developed the potential of fluorescence temperature sensing in shocked materials. In particular, the results demonstrated that pressure has no substantive significance of the structure of fluorescence spectrum [22]. Based on these interesting phenomena, we try to observe the temperature dependence of the fluorescence spectrum of anthracene crystals and find an applicable physical quantity for fluorescence temperature sensing, which was expected to have the advantage of high sensitivity and low random uncertainty.

## 2. Experimental

To establish a new optical temperature sensing method for shocked materials, the steady state fluorescence spectral measurement system at varying temperature was established, as shown in Fig. 1 (a). The anthracene powder was purchased from Sigma-Aldrich at >99% purity. A SuperK EXTREME super-continuum laser (NKT Photonic), whose wavelength can be tunable in the range from 420nm to 680nm, was used as the excitation source. The excitation wavelength was 440nm in this work. After propagating through a dichroic mirror $M_2$, the pump light was focused on the round specimen chamber of an Air Products Displex closed-cycle helium cryostat by the focusing lens $L_1$, which provided continuously variable temperature from 300K to 500K. The Rayleigh scattered light removed by an optical filter $F_1$. Finally, another focusing lens $L_2$ was used to focusing the scattered light into the optical fiber. The fluorescence spectrum was recorded by using a half meter spectrometer with a CCD camera (Andor) as a detector. The spectral resolution of the spectrometer is 0.179nm. The time-resolved dynamics of different fluorescence transitions were detected by using Time-Correlated Single Photon Counting (TCSPC) systems (resolving capability 180ps, PMC-100-1, Becker & Hickl GmbH).

To obtain an unbroken fluorescence spectrum structure of anthracene crystals, the JB 450 was used as the optical filter and the fluorescence spectrum of anthracene at normal pressure and temperature was observed, as shown in Fig. 1 (b). The fluorescence spectrum showed five transitions (denoted as 2-0 and 1-n, here n = 0-3, respectively) from excited state to the ground state. The characteristic of those transitions were obtained by multi Gaussian peaks fitting. The weak peak near the 2-0 transition was originated from the incomplete filter effect of the JB 450 optical filter and be ignored in the following analysis. The fluorescence spectrum of the broad 1-3 transition can't be precisely ascertained and also be not taken into account in our study.

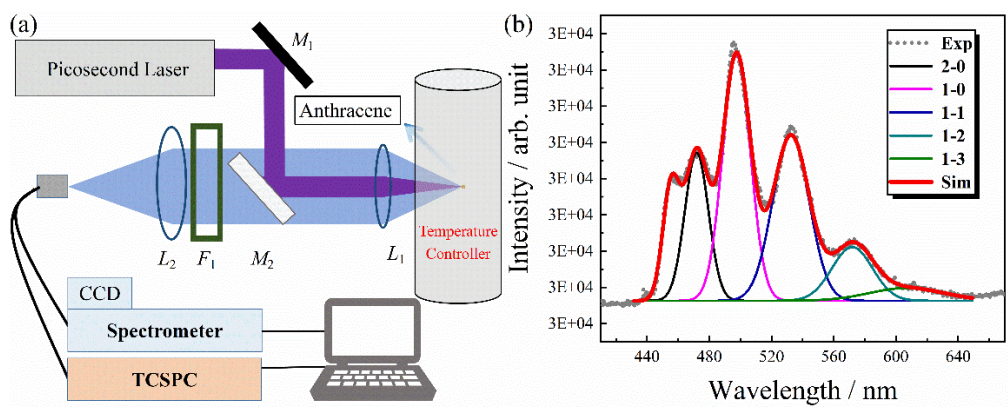

Fig. 1 (a) The experimental device diagram. $L_1$, $L_2$: lens, $F_1$: optical filter, $M_1$: reflector mirror, $M_2$: dichroic mirror; (b) the fluorescence spectrum of anthracene crystals at normal pressure and temperature.

## 3. Results and discussion

### 3.1 Temperature dependence of fluorescence spectrum

In order to establish a fluorescence temperature sensing method, the fluorescence spectrum of anthracene crystals at varying temperature in the range from 300K to 500K was recorded in the following experiment, as shown in Fig. 2. It is obvious that there was a significant structural change in the fluorescence spectrum of anthracene crystals with increasing temperature. For letting the analysis more detailed, the fluorescence spectrum of anthracene crystals at varying temperature was fitting by multi Gaussian peaks and the characteristic of different transitions obtained. The impressive phenomenon in the fluorescence spectrum of anthracene crystals was the remarkable increasing of the intensity of 2-0 transitions, which was at a wavelength of about 470 nm. To date, numerous researches have explored the mechanism of the fluorescence emission increasing with temperature increasing [23-25]. It has been widely acknowledged that the thermally activated transition from singlet state to triplet state and the reabsorption were both responsible for the temperature induced fluorescence emission strengthen phenomenon [23-25]. Note that the temperature sensitivity of the central wavelength and the full width at half maximum (FWHM) of fluorescence spectrum are too lower to open the possibilities for temperature sensing, the temperature dependence of these two physical quantities were omitted here for the sake of brevity. Here, we focused on the temperature dependence of the intensities of different transitions in anthracene crystals.

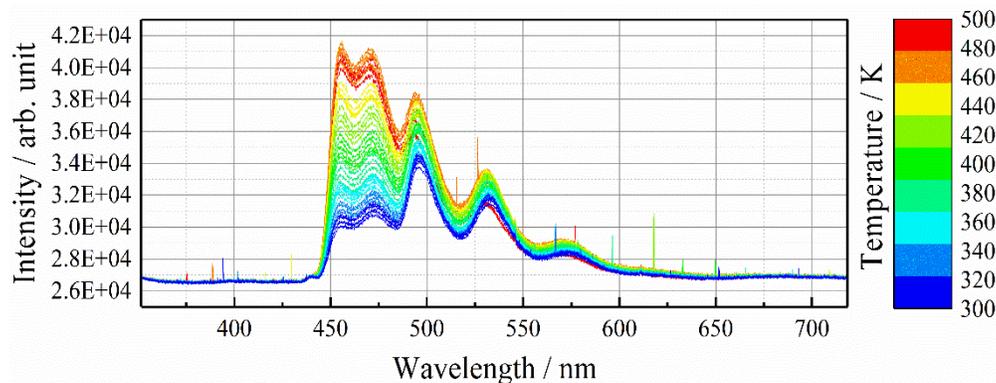

Fig. 2. Temperature dependence of fluorescence spectrum of anthracene crystals.

*3.2 Temperature dependence of intensity for different transitions*

In this section, the temperature dependence of the intensities for different transitions, (2-0 and 1-n, n=0-2) were shown in Fig. 3 (a). It is apparent that the 1-0, 1-1 and 1-2 transitions all showed same evolution tendency with the increasing temperature. What is striking about the figure is that the intensity of 2-0 transition was very sensitive to temperature and showed a remarkable rising tendency until the temperature reached 470K. It's worth noting that the intensity of 2-0 transition at 470K was three times than the value at 300K and surpassed the intensity of 1-0 transition at 420K. In addition, the growth curves for all transitions was ended around 470K, which was attribute to the solid-liquid phase transition of anthracene crystals. We note that the absolute intensity was influenced by many factors, such as the excitation energy, the associated settings of the CCD and so on. Hence, the absolute intensity can't be used directly for fluorescence temperature sensing.

Note that the intensity ratio of two energy-close fluorescence emission was rely on the populations of the emission states, the intensity ratio was widely used for fluorescence temperature sensing [26-28]. Considering the intensities of 1-0, 1-1 and 1-2 transitions showed same increasing tendency, we defined a new physical quantity named intensity ratio, $\gamma_n=I_{2-0}/I_{1-n}$, n=0,1,2 for further analysis. Here, $I_{2-0}$ is the intensity of 2-0 transition and $I_{1-n}$ is the intensity of the 1-n transition respectively. Then, the temperature dependence of different intensity ratios $\gamma_n$ were calculated and shown in Fig. 3 (b), (c) and (d), which correspond to $\gamma_0$, $\gamma_1$ and $\gamma_2$ respectively. For the sake of discussion, we name the fluorescence temperature sensing methods based on those different intensity ratios $\gamma_0$, $\gamma_1$ and $\gamma_2$ as methods $M_0$, $M_1$ and $M_2$ respectively. Interestingly, these three different intensity ratios $\gamma_n$ all showed perfect exponential increasing tendency with increasing temperature. In particular, the exponential increasing tendency still last even the solid-liquid phase transition of anthracene crystals was begun. The detailed exponential fitting information for these three different intensity ratios were shown in the Table 1. This phenomenon demonstrated that the solid-liquid phase transition of anthracene crystals was unable to influence the temperature dependence of intensity ratios. Hence, the fluorescent temperature sensing method based on the intensity ratio will not be limited by the solid-liquid phase transition.

Table 1. The detailed fitting information for $\gamma_1$, $\gamma_2$ and $\gamma_3$.

|  | $y_0$ | | $A_1$ | | $t_1$ | | R-Square |
|---|---|---|---|---|---|---|---|
|  | Value | Standard Error | Value | Standard Error | Value | Standard Error |  |
| $\gamma_1$ | 0.19051 | 0.03841 | 0.06739 | 0.0117 | 165.55593 | 8.21277 | 0.99751 |
| $\gamma_2$ | 0.53165 | 0.05532 | 0.02425 | 0.0064 | 109.46995 | 5.94932 | 0.99371 |
| $\gamma_3$ | 2.5113 | 0.23702 | 0.00673 | 0.00472 | 71.04864 | 7.01212 | 0.96059 |

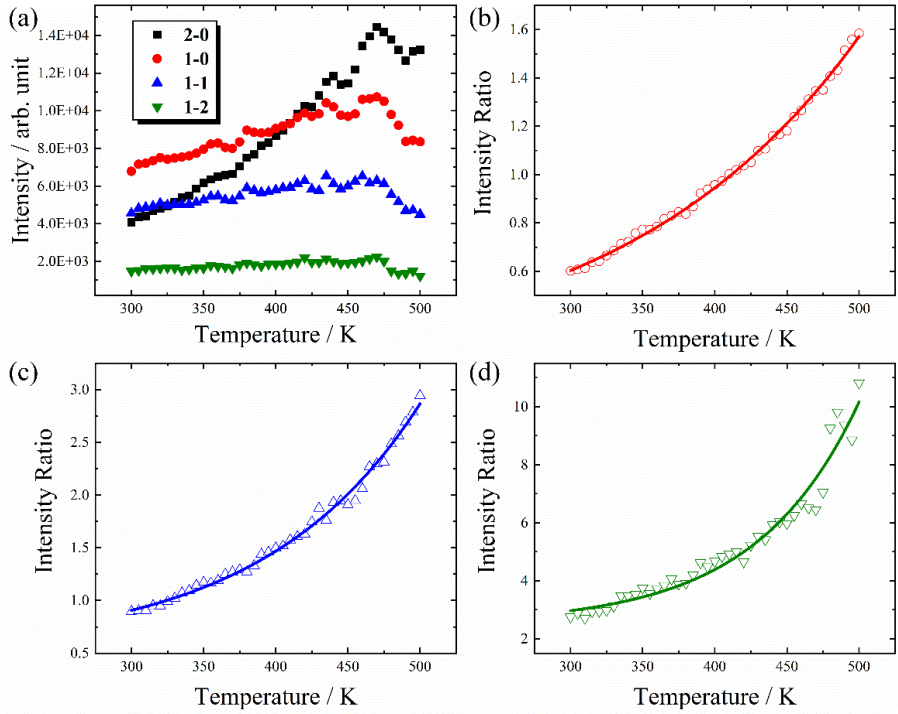

Fig. 3 (a) The intensity of different transitions at varying temperature; the temperature dependence of intensity ratios (b) $\gamma_0$, (c) $\gamma_1$, (d) $\gamma_2$.

*3.3 The performance of different fluorescence temperature sensing methods*

In this part, to evaluate the property of the fluorescence temperature sensing method, the sensitivity $\eta_n = d\gamma_n/\gamma_n dT$ and random uncertainty $\Delta T_n = [(\Delta I_{2-0}/I_{2-0})^2 + (\Delta I_{1-n}/I_{1-n})^2]^{1/2}/\eta_n$ was analyzed respectively. Here, the $\Delta I_{2-0}$ and $\Delta I_{1-n}$ was the random uncertainty for 2-0 transition and 1-n transition. Firstly, we calculated the sensitivity at varying temperature for different methods $M_0$ (red), $M_1$ (blue), $M_2$ (green), and plotted in the Fig. 4 (a). It is obvious that all the sensitivities are independent of temperature. It can be seen from the figure that method $M_2$ was the most sensitive fluorescence temperature sensing method, and the method $M_1$ ranked second. What stands out is the sensitivity for method $M_2$ is about 0.014 $K^{-1}$, which was over two times than the value of method $M_0$. Secondly, the random uncertainty $\Delta T_n$ at varying temperature for these three different methods were calculated, as shown in Fig. 4 (b). Unfortunately, the method $M_2$ showed an unsatisfactory performance that the mean random uncertainty was around 6.5 K and the maximum value reached up to 10.8K. In spite of possessing the advantage of high sensitivity, the method $M_2$ is not meeting the recommended level for accurate temperature sensing because of the higher random uncertainty. Interestingly, the random uncertainty for methods $M_0$ and $M_1$ were both showed a slightly fluctuated around 1.0K. Even the random uncertainty of method $M_0$ and method $M_1$ were perform at the same level, the method $M_1$ was more appropriate for fluorescence temperature sensing with the advantage of higher temperature sensitivity. In general, the method $M_1$ is the best comprehensive performance method for fluorescence temperature sensing and the remaining methods have own limitation, that lower sensitivity for method $M_0$ and higher random uncertainty for method $M_2$ respectively. If the precision requirement of temperature sensing is not strict, the method $M_2$ may be performing well under this situation.

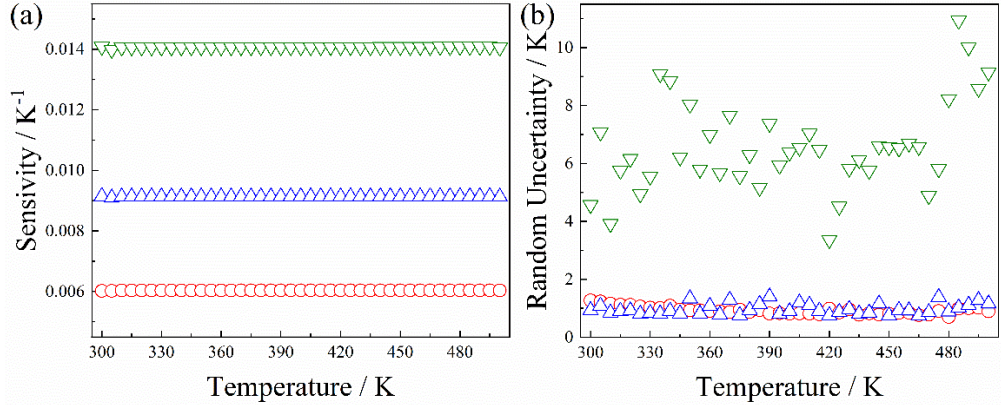

Fig. 4 (a) The sensitivity $\eta_n$ and (b) the random uncertainty $\Delta T_n$ at varying temperature for methods: $M_0$ (Red), $M_1$ (Blue) and $M_2$ (Green).

Based on the former analysis, the method $M_1$ based on the intensity ratio $\gamma_1$ was selected as a representative method of our study for the following comparison. The property of several typical fluorescent temperature sensing methods was summarized in Table 2. Due to the strong fluorescence intensity, the dye molecule was widely used for the optical temperature sensing. The organic dye 1, 3-bis propane has the advantage of high sensitivity and low random uncertainty but in a narrow temperature range [29]. However, the sensitivity and random uncertainty of anthracene probe in our work can reach the same level with some inorganic dyes, such as: YAG:Ce or YAG:Dy [30, 31]. In addition, it can be seen that the sensitivity and random uncertainty of our method was able to comparable to the traditional dual color Stokes fluorescence intensity ratio technique, which based on the $Mg_4FGeO_6$:Mn [32]. When compared with CdSe quantum dot, the widely temperature sensing range of our method can attract more attention for luminescence thermometry [33]. In general, the anthracene probe was suitable for accurate temperature sensing in the temperature range 300K to 500K.

Table 2. Excitation Wavelength ($\lambda_{ex}$), Temperature Range (TR), Sensitivity ($\eta$) and random uncertainty ($\Delta T$) for different fluorescence temperature sensing methods.

| Probe | $\lambda_{ex}$ (nm) | TR (K) | $\eta$ ($10^{-4}K^{-1}$) | $\Delta T$ (K) | Reference |
|---|---|---|---|---|---|
| Anthracene | 440 | 300-500 | 9 | 1.0 | This work |
| 1,3-bis propane | 325 | 303-413 | 140 | 0.3 | [29] |
| YAG:Ce | 527 | 300-800 | 7 | 4.0 | [30] |
| YAG:Dy | 355 | 300-1673 | 11 | 9-50 | [31] |
| $Mg_4FGeO_6$:Mn | 266 | 300-800 | 3 | 5-22 | [32] |
| CdSe/ZnS | 900 | 315-330 | 3 | 2.0 | [33] |

*3.4 The mechanism for temperature sensing based on intensity ratio*

In the above section, we find the intensity of 2-0 transition showed a remarkable increasing tendency with the increasing temperature. Then, we explored the possibility of fluorescence temperature sensing based on intensity ratio and got a satisfactory result. In this section, we have tried to clarify the origin of the remarkable strengthen of 2-0 transition with increasing temperature.

Firstly, researchers has confirmed that there is little or no internal conversion from excited state $S_1$ to the ground state $S_0$ of anthracene for aromatic hydrocarbons including anthracene, which indicated that the radiative transition fluorescence with high probability [34, 35]. Note the probability of non-radiative intersystem crossing has positive correlation with the radiative transition probability, it was more likely to observe the non-radiative intersystem crossing phenomenon at anthracene crystals [35]. In addition, this intersystem crossing was usually considered to be the first-order spin-orbit coupling between electronic states [36]. Hence, different electron distribution will lead to different efficiency intersystem crossing. Meanwhile, the $\pi$ electron maxima at the atoms of the first excited $S_1$ ($^1L_\alpha$) state have more tendency to undergo intersystem crossing [35, 37].

Secondly, theory has predicted that there are usually at least three triplet ($\pi, \pi^*$) state predicted lie within 3000cm$^{-1}$ of the lowest excited singlet state $S_1$ for anthracene [35, 38]. The large Frank-Condon factor limited the channel efficiency of intersystem crossing from $S_1$ to lowest excited triplet state $T_1$ [39, 40]. In particular, the rate of directly intersystem crossing for this case is independent of temperature [41]. In fact, the most effective intersystem crossing from $S_1$ to $T_1$ pathway is that from $S_1$ via $T_2$ to $T_1$ [42]. Previous research have shown the existence of a second excited triplet state, which was laying below the first excited singlet state in solution but above the first excited singlet state in the crystals [25, 39, 43]. In addition, the reported energy gap was between the first excited singlet state $S_1$ and the second excited triplet state $T_2$ were about 800cm$^{-1}$ [44], 860cm$^{-1}$ [25] and 1200cm$^{-1}$ [39] respectively. What's more, there is considerable evidence of intersystem crossing from singlet state $S_1$ to triplet state $T_2$ was thermally activated [44], which was the reason for the intensity ratio be not sensitive to pressure.

Finally, based on the former research results and the observed experimental results, we proposed that the observed 2-0 transition was emitted from the second excited triplet state $T_2$, as shown in Fig. 5(a). Considering the energy gap was sensitive to the environment the state of matter [47], the energy gap in our study was about 1095cm$^{-1}$ and the deviation was reasonable. Due to the energy difference $\Delta E$ was at the range from 200 to 2000cm$^{-1}$, the two emitting levels could be considered "thermally coupled" and assumed as a Boltzmann thermal equilibrium model. Hence, the population of the two emitting states $S_1$ and $T_2$ could be related by [46, 47]:

$$N_0 = N \left( \frac{g_0}{g} \right) \exp\left( \frac{-\Delta E}{k_B T} \right) \tag{1}$$

Here, $g$ and $g_0$ are the degeneracies of $S_1$ and $T_2$ states, $N$ and $N_0$ are the population of $S_1$ and $T_2$ states. Hence, the intensity of 1-0, 1-1 and 1-2 transitions can be expressed as: $I_{1-n}=\hbar\omega_{1-n}A_{1-n}N$, respectively. The $A_{1-n}$ and $\omega_{1-n}$ are the total spontaneous emission rates and the corresponding frequencies for the 1-n transitions respectively. However, the intensity of 2-0 transition, which was emitted from the second excited triplet state $T_2$, can be expressed as: $I_{2-0}=\hbar\omega_{2-0}A_{2-0}N_0$. Similarly, the $A_{2-0}$ and $\omega_{2-0}$ are the corresponding total spontaneous emission rate and frequency. In the end, we can obtain the mathematical expression of the intensity ratio as:

$$\gamma_n = \frac{\hbar\omega_{2-0}A_{2-0}N_0}{\hbar\omega_{1-n}A_{1-n}N} = \frac{\omega_{2-0}A_{2-0}g_0}{\omega_{1-n}A_{1-n}g} \exp\left( \frac{-\Delta E}{k_B T} \right) \tag{2}$$

It can see from the formula (2) that the intensity ratio will exponentially increasing with temperature. Meanwhile, researches had reported that the singlet-triplet intersystem crossing rate constant $K_{TM}$ increases with the rising temperature according to the relationship $K_{TM}=K_{TM}^0 + A_{TM} \exp(-E_{IS}/k_B T)$ [45]. Here $E_{IS}$ represents an activation energy for intersystem crossing, which is equal to the $\Delta E$ energy difference in our study, $A_{TM}$ is the proportionality coefficient and $K_{TM}^0$ is the singlet-triplet intersystem crossing rate at initial temperature. This equation has demonstrated that there was an exponential relationship between the population of the two emitting states $S_1$ and $T_2$. Essentially, our theoretical result that equation (1) was coincide with the reported research results.

In spite of the analytical results showed an excellent agreement with the experimental results, we still need some experimental evidences to support our explanation. To demonstrate that the observed 2-0 transition was originated from another triplet state rather than the first excited singlet state $S_1$, we observed the lifetime of different transitions at varying temperature range from 300K to 500K, respectively, as shown in Fig. 5 (b), (c), (d) and (e). It was worth noting that the lifetime of 1-0, 1-1 and 1-2 transitions all showed an increasing tendency with the increasing temperature. In particular, the lifetime variation tendency of 1-1 transitions was very similar to the 1-2 transition. However, the lifetime variation tendency of 1-0 transition showed a little difference with respect to the 1-1 and 1-2 transitions. This difference can be ascribed to the overlap of the 2-0 and 1-0 transitions, which lead the measured spectrum contain the contribution of 2-0 transition. Based on this fact, it is reasonable to draw a conclusion that the temperature should have the same effect on the fluorescent lifetime of different transitions, as long as these transitions were originated from the same emitting state. Meanwhile, the temperature dependence of the lifetime of 2-0 transition was found to be negligible, which was consistent with other research results [48]. This unusual phenomenon indicated that the origin of the observed 2-0 transition was another emitting state, which was belong to triplet state.

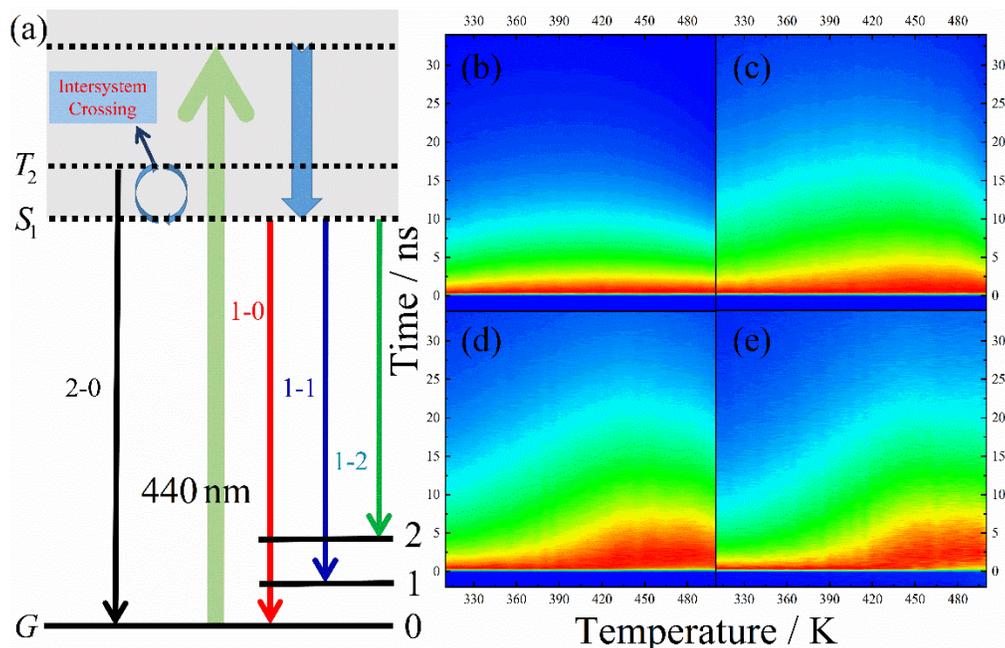

Fig. 5 (a) The energy level diagram of anthracene crystals; the temperature dependence of fluorescent lifetime for (b) 2-0, (c) 1-0, (d) 1-1 and (e) 1-2 transitions.

## 4. Summary

In this work, we observed the steady-state fluorescence spectrum and the time-resolved dynamics of different transitions of anthracene crystals at varying temperature range from 300K to 500K. By multi-peaks Gaussian fitting, the fluorescence spectrum of different components, that 2-0, 1-0, 1-1 and 1-2 transitions, were characterized respectively. Fortunately, even the solid-liquid phase transition begun, the intensity ratio still showed a perfect exponential increasing curve in the temperature range from 300K to 500K. The fluorescence temperature sensing method $M_1$ based on intensity ratio $γ_1$ was proved to be the best comprehensive performance method after the evaluation of sensitivity $η$ and random uncertainty $ΔT$. In addition, theoretical analysis and experimental results clarified the unusual intensity increasing of 2-0 transition was emitted from the second excited triplet state $T_2$, which was above the first excited singlet state $S_1$. Particularly, the exponential temperature dependence of intensity ratio was proved to be the consequence of thermally activated singlet-triplet intersystem crossing. Note that the fluorescence spectrum of anthracene single crystal was insensitivity to the pressure, the methods reported in this paper can develop the potential of accurate temperature sensing in shocked materials and open new perspectives in the fluorescence research of aromatic molecules.

## Acknowledgments

This work is supported by the National Natural Science Foundation of China 51372057 and Science Challenge Project (Grant No. TZ2016001).